\documentclass[lettersize,journal]{IEEEtran}
\usepackage{amsmath,amsfonts}
\usepackage{array}
\usepackage[caption=false,font=normalsize,labelfont=sf,textfont=sf]{subfig}
\usepackage{textcomp}
\usepackage{stfloats}
\usepackage{url}
\usepackage{verbatim}
\usepackage{graphicx}
\usepackage{bm}
\usepackage{cite}

\usepackage{times}
\usepackage{fancyhdr,graphicx,amssymb}
\usepackage{algorithm} 
\usepackage{algorithmic} 

\usepackage{tabularx} 
\usepackage[table]{xcolor} 
\usepackage{booktabs} 
\usepackage{multirow}
\usepackage{array} 

\def\BibTeX{{\rm B\kern-.05em{\sc i\kern-.025em b}\kern-.08em
    T\kern-.1667em\lower.7ex\hbox{E}\kern-.125emX}}
\usepackage{balance}
\begin{document}
\bstctlcite{BSTcontrol}
\title{FAST: Flexible and Adaptive Semantic Transmission for  Resource-constrained Multi-user Generative Semantic Communication}
\author{Yiru Wang, Wanting Yang, Fangli Mou, Zehui Xiong,~\IEEEmembership{Senior~Member,~IEEE, }

Zide Fan,~\IEEEmembership{Member,~IEEE, }Shiwen Mao,~\IEEEmembership{Fellow,~IEEE, }Tony Q. S. Quek,~\IEEEmembership{Fellow,~IEEE, }

\thanks{
Yiru Wang, Fangli Mou and Zide Fan are with the Aerospace Information Research Institute, Chinese Academy of Sciences, Beijing 100190, China, and also with the Key Laboratory of Target Cognition and Application Technology, Aerospace Information Research Institute, Chinese Academy of Sciences, Beijing, China(wangyiru@aircas.ac.cn; moufangli@aircas.ac.cn; fanzd@aircas.ac.cn);
Wanting Yang is with SUTD Wireless Innovation Center, Singapore University of Technology and Design, Singapore (e-mail: wanting\_yang@sutd.edu.sg);
Zehui Xiong is with the School of Electronics, Electrical Engineering and Computer Science (EEECS), Queen's University Belfast, Belfast, BT7 1NN, U.K. (z.xiong@qub.ac.uk);
Shiwen Mao is with the Department of Electrical and Computer Engineering, Auburn University, Auburn 36830, USA (e-mail: smao@ieee.org);
Tony Q. S. Quek is with the Pillar of Information Systems Technology and Design, Singapore University of Technology and Design, Singapore (tonyquek@sutd.edu.sg).
}}

\markboth{Journal of \LaTeX\ Class Files,~Vol.~18, No.~9, September~2020}%
{How to Use the IEEEtran \LaTeX \ Templates}

\maketitle

\begin{abstract}
The rapid advancement of generative artificial intelligence has spurred innovative approaches to semantic communication, giving rise to a new paradigm known as generative semantic communication (GSC). The integration of flexible cross-modal semantic extraction with generative capability–driven semantic inference substantially enhances semantic compression efficiency, demonstrating significant promise under communication resource constraints. Nonetheless, the stringent dependence on high computational power and the resulting latency continue to present major challenges, thereby limiting the feasibility of large-scale deployment. To address these challenges, we propose a novel GSC framework named FAST, which stands for \underline{f}lexible and \underline{a}daptive \underline{s}emantic \underline{t}ransmission. To accommodate limited computational resources,  we propose a sequential semantic extraction method, where a temporal prompt engineering module orchestrates the distillation and transmission of key semantic units. Correspondingly, we introduce a sequential conditional denoising module at the receiver, which adapts the diffusion-based reconstruction to the progressively received input. To enhance overall task performance in multi-user semantic transmission, we propose a semantic-aware resource allocation method that optimizes bandwidth dynamically based on a joint consideration of semantic dependencies, user-level task priorities, and instantaneous channel conditions. Extensive experiments demonstrate that the proposed architecture achieves system precision comparable to conventional GSC systems while significantly reducing transmission latency and improving overall efficiency. These results confirm its enhanced potential for deployment in multi-user GSC scenarios with stringent communication and computational constraints.


\end{abstract}

\section{Introduction}
\IEEEPARstart{T} {he} imminent advent of sixth-generation (6G) networks is anticipated to experience an unparalleled increase in data traffic, fueled by the proliferation of cutting-edge applications such as ultra-high-definition video streaming, vehicular networks, and extensive Internet of Things (IoT) deployments. Concurrently, traditional communication systems are approaching their limit, and the scarcity of available spectrum resources is becoming more pronounced. These factors collectively present substantial challenges in ensuring the quality of user experiences for these advanced services \cite{agentdriven}. Recognizing the capacity of generative artificial intelligence (GAI) for modal transformation and its exceptional ability to produce high-quality content across various domains, generative semantic communication (GSC) presents a transformative solution. In GSC systems, task-related semantic information is distilled in the form of a \textit{prompt}, which is transmitted to the receiver to guide the GAI models in reconstruction \cite{harnessing}. This approach reduces the required data payload and enhances the overall efficiency of information exchange. Recent research validates the effectiveness of using image captions \cite{xia}, segmentation maps \cite{monitor}, and skeleton maps \cite{skeleton} as prompts to reduce the transmission payload in image transmission scenarios.


However, conventional GSC systems \cite{xia, monitor, skeleton} lack the adaptability for complex tasks like digital twins and virtual reality games, which are characterized by heterogeneous data sources, diverse user task criticality, and time-varying channel conditions from mobility and obstruction. For instance, a digital twin system integrates data from multiple physical sensors, with each source representing distinct semantic dimensions—from entity states and environmental interactions to system dynamics. Concurrently, system users (e.g., automated guided vehicles or inspection workers) are often mobile while also exhibiting diverse task-specific demands for critical functions such as synchronization, prediction, and optimization, each imposing distinct requirements on latency, reliability, and data fidelity. The interplay of these challenges is further compounded by the inherent computational and communication resource constraints at the network edge, creating a critical bottleneck for conventional transmission designs.

To overcome these limitations, we introduce a \underline{f}lexible and \underline{a}daptive \underline{s}emantic \underline{t}ransmission (FAST) framework for computation and communication resource-constrained multi-user GSC scenarios that efficiently handles heterogeneous data and adaptively prioritizes tasks based on their criticality. The contributions of this work are outlined as follows:
\begin{itemize}
    \item To accommodate limited computational resources, we propose a sequential semantic extraction method. In this approach, a temporal prompt engineering module governs the execution order of extraction models and distills key semantic units for transmission in a sequential manner. Specifically, this module employs a reinforcement learning (RL) agent to discern the intrinsic sequential dependencies among the semantic units and adapt their extraction sequences according to the task. To enhance learning efficiency, we incorporate an invalid action masking technique, which prunes the action space by excluding choices that do not positively contribute to the performance reward.
    \item At the user end, we introduce a sequential conditional denoising module that adapts the diffusion-based reconstruction process to the sequential transmission mechanism. Specifically, we devise a semantic difference calculation subunit that accurately identifies the areas within the noise space targeted by newly arrived semantic units. This identified difference is then applied to the predicted conditional noise, enhancing the guidance effectiveness of late-arriving semantic units and alleviating the stringent immediacy requirement.
    \item To enhance overall task performance in multi-user semantic transmission, we develop a semantic-aware resource allocation module. This module performs dynamic bandwidth allocation by synthesizing three critical factors: semantic dependencies among data units, user-level task priorities, and instantaneous channel conditions.
    \item In our experimental evaluation, we separately assess the effectiveness of the aforementioned three components. The efficacy of these enhancements is conclusively demonstrated through both quantitative analyses and visual evidence. Building upon this validation, we synergistically integrate these modules, culminating in the establishment of the FAST framework. Extensive experiments demonstrate that the proposed architecture achieves task precision comparable to conventional GSC systems, while significantly reducing transmission latency and improving overall system efficiency.
\end{itemize}

The rest of this paper is organized as follows. The background and related work on GSC are provided in Section \ref{Related Work}. The overall architecture of the proposed FAST framework is presented in Section \ref{System Model}, and its three core components are described in Section \ref{components}. The experimental evaluations are conducted in Section \ref{Experiments}, followed by the conclusions of the study in Section \ref{Conclusion}.

\section{Background and Related Work} \label{Related Work}
In this section, we review the background and related work on GSC. We first introduce the foundational components of GSC, which encompass semantic communication principles, conditional diffusion models, and prompt engineering techniques. Subsequently, we critically review relevant studies in GSC, and compare them with our proposed FAST framework.

\subsection{Fundamentals of GSC} 

\textit{1) Semantic Communications:} Semantic communication has emerged as a paradigm shift that moves beyond the conventional goal of transmitting bits accurately toward the precise delivery of meaning \cite{Luo, Getu2024, Guo2025, xie2021}. Unlike traditional systems that focus primarily on signal-level fidelity, semantic communication seeks to ensure the successful execution of a downstream task, such as classification \cite{classification1, classification2}, reconstruction \cite{speech, image1, image2}, or decision-making \cite{decision}. By discarding semantically redundant information, this approach demonstrates significant potential for extreme data compression and bandwidth savings, particularly in resource-constrained scenarios. 

Existing semantic communication frameworks typically leverage a joint source-channel coding (JSCC) architecture, implemented via end-to-end trained deep neural networks, to extract and transmit task-relevant semantic features. Xie et al. \cite{xie2021deep} proposed a Transformer-based semantic communication system for text transmission. In the proposed framework, a JSCC framework was designed to extract the semantic information from texts and cope with channel noise and semantic distortion. Yang et al. \cite{WITT} redesigned the vision Transformer as a backbone for image transmission. This JSCC framework was optimized to extract long-range information from images and consider the effect of the wireless channel. 

Traditional SemCom faces two principal challenges: its reliance on end-to-end models results in poor interpretability, making debugging costly; furthermore, the framework exhibits limited extensibility, requiring major redesigns to handle multi-modal tasks and thus hindering its broad applicability.

\textit{2) Conditional Diffusion Models:} Conditional diffusion models have emerged as a powerful class of generative models capable of producing high-quality samples across a range of data modalities \cite{DDPM,DDIM,lipman2022flow, liu2022flow}. These models iteratively refine a distribution of noise to generate samples, conditioned on external information to guide the generation process towards desired characteristics. To exert a high degree of control over the generation process, a diverse array of conditions can be employed. Methods that leverage text for guidance have demonstrated remarkable capabilities across various media types \cite{text1,text2}. For delineating the contours of image objects, several studies have employed segmentation masks as conditioning inputs \cite{seg1}. Additionally, the potential for manipulation through other modalities such as keypoints and depth maps has been explored, further expanding the versatility of control \cite{other1,other2}.

To address the computational challenges of deploying diffusion models on resource-constrained devices, Du et al. \cite{du2023exploring} proposed a distributed denoising framework. In this approach, the initial denoising steps are offloaded to an edge server and shared across multiple users. The resulting intermediate latent representations are then transmitted to individual user devices, which complete the remaining task-specific denoising steps. This strategy not only improves computational efficiency but also enhances privacy by allowing users to retain control over their final output. However, a key limitation is that user-specific prompts are introduced only after the shared denoising process. This delayed induction weakens their influence on the generated content, potentially leading to suboptimal reconstruction quality.

To address this limitation, Du et al. \cite{User-centric} enforced a constraint whereby the shared initial denoising steps were executed exclusively among users possessing semantically similar prompts. In a concurrent approach, Xie et al. \cite{gaochang} proposed grouping users based on task similarity and reducing divergence among the late-induced user-specific prompts, thereby enhancing the reconstruction performance at a group level. However, this method does not fundamentally address the root cause of performance degradation—the inherently limited influence of any prompt introduced after critical early denoising steps. This core limitation persists and necessitates more effective solutions.

\textit{3) Prompt Engineering:} Recent research in this field indicates that advanced models like DALL-E \cite{dall-e}, Stable Diffusion \cite{stable}, and GPT-4 \cite{gpt} respond variably to subtle nuances in language, where even minor modifications in the prompt can lead to significantly different outputs. In light of this phenomenon, prompt engineering emerges as a crucial technique for boosting the performance of GAI models. Within this field, several methods are prominently utilized. For example, Brown et. al \cite{engineering1} proposed a keyword-based prompting approach that incorporates specific keywords or phrases to guide the model's attention, ensuring alignment with the desired topic or context. Gao et. al \cite{engineering2} and Liu et. al \cite{prompt_engineering} developed context-rich prompting which provides detailed scenarios or background information to frame the response accurately, making the generated output more contextually appropriate and relevant. Feng et. al \cite{engineering5} proposed iterative refinement which adjusts prompts based on feedback and results, allowing for continuous improvement in the model’s performance.

Nonetheless, given that the temporally segmented denoising approach has only recently been introduced in references \cite{du2023exploring, User-centric, gaochang}, the implementation of prompt engineering in a time-sequential manner remains areas yet to be explored.

\subsection{Review of the Existing Studies in GSC} 

To enhance the performance of GSC, a series of transmission designs have been developed. These approaches can be broadly categorized into those tailored for single-user and multi-user communication scenarios.

\textit{1) Single-user GSC Scenario:} Within single-user scenarios, research has shown that GAI holds significant promise for enhancing key system performance metrics of semantic communications.

Grassucci et al. \cite{grassucci2023generative} proposed a diffusion-guided semantic communication framework that capitalizes on the powerful generative capability of diffusion models in synthesizing multimedia content. This approach achieved a significant reduction in bandwidth usage while effectively preserving semantic integrity. Yang et al. \cite{Yang2024WCNC} developed a semantic change-driven generative semantic communication framework for remote monitoring applications. Their design incorporated a modular semantic encoder equipped with a value-of-information-based sampling function, along with a conditional denoising diffusion probabilistic model acting as the semantic decoder to accurately reconstruct remote scenes. Experimental results validated that the proposed encoder-decoder architecture exhibited significant potential for minimizing energy consumption. In their subsequent work \cite{agentdriven}, they designed an agent-assisted semantic encoder with cross-modality capabilities, which further reduced the transmission workload and enhances energy efficiency. Guo et. al \cite{guo2025diffusion} designed an advanced variational auto-encoder-based compression method for bandwidth-constrained GSC systems. To achieve low-latency and high-quality video transmission over bandwidth-constrained noisy wireless channels, Li et al. \cite{li2025goal} proposed a diffusion-based goal-oriented semantic communication framework. The framework incorporated a paired semantic encoder-decoder to efficiently extract and reconstruct video information, thereby reducing transmitted data size, along with a diffusion model-based denoiser to mitigate channel-induced noise. 

However, focusing solely on single-user optimization leads to a sub-optimal system-level performance, as it ignores the potential for coordinated transmission and scheduling management. Therefore, advancing multi-user GSC is imperative to unlock the full potential of semantic-aware networks and transition from isolated links to an intelligent system.

\textit{2) Multi-user GSC Scenario:} The extension of GSC from single-user to multi-user architectures is not merely a scalability enhancement but a critical step toward practical deployment. The multi-user GSC scenarios involve multiple agents or devices with distinct tasks, data semantics, and quality-of-service requirements, all competing for shared and often severely limited communication and computational resources. Therefore, a well-designed transmission strategy is essential to further enhance the flexibility and adaptability of the network.

Grassucci et al. \cite{grassucci2024rethinking} proposed a channel allocation strategy based on the generative capability of diffusion models and utilized these models to reconstruct missing information at the receivers. This strategy enabled more efficient bandwidth allocation across multiple users. To address diverse yet overlapping user demands, Lu et. al \cite{lu2025multi} proposed a multi-user GSC-enhanced intent-aware semantic-splitting multiple access framework. Within this framework, the common semantic information was designated as a one-hot semantic map that was broadcast to all users, while the private semantic information was delivered as personalized text for each user. Wang et. al \cite{Wang2025SFMA} investigated multi-user GSC system for video transmission, which comprised a base station and paired users. The base station generated and combined semantic information of several frames simultaneously requested by paired users into a signal. Users recovered their frames based on a GAI-based video frame interpolation model. To optimize transmission rates and temporal gaps between simultaneously transmitted frames, an optimization problem was formulated to maximize the system sum rate while minimizing temporal gaps. To overcome substantial bandwidth consumption and potential transmission failures of mobile AI-generated content service, Liu et. al \cite{liu2024cross} applied cross-modal GSC framework and formulated a joint semantic encoding and prompt engineering problem to optimize the bandwidth allocation among users. Ren et. al \cite{ren2025generative} developed a multi-user GSC framework leveraging pretrained Multi-modal/Vision Language Models and formulated a joint optimization problem of prompt generation offloading, communication and computation resource allocation to minimize the latency and maximize the resulting semantic quality.

The transition to multi-user semantic communication poses a significant challenge, primarily due to user task heterogeneity and stringent resource constraints—a critical gap that remains largely unaddressed in the existing literature. To situate our contribution within this context, we systematically outline the aforementioned frameworks and present a comparative analysis with our proposed FAST method in Table \ref{table}.

\begin{table*}[htbp]
\centering
\caption{A Summary and Comparison of Existing GSC Studies and Our Work}
\vspace{-0.2cm}   
\begin{tabular}{c}
\includegraphics[width=18cm]{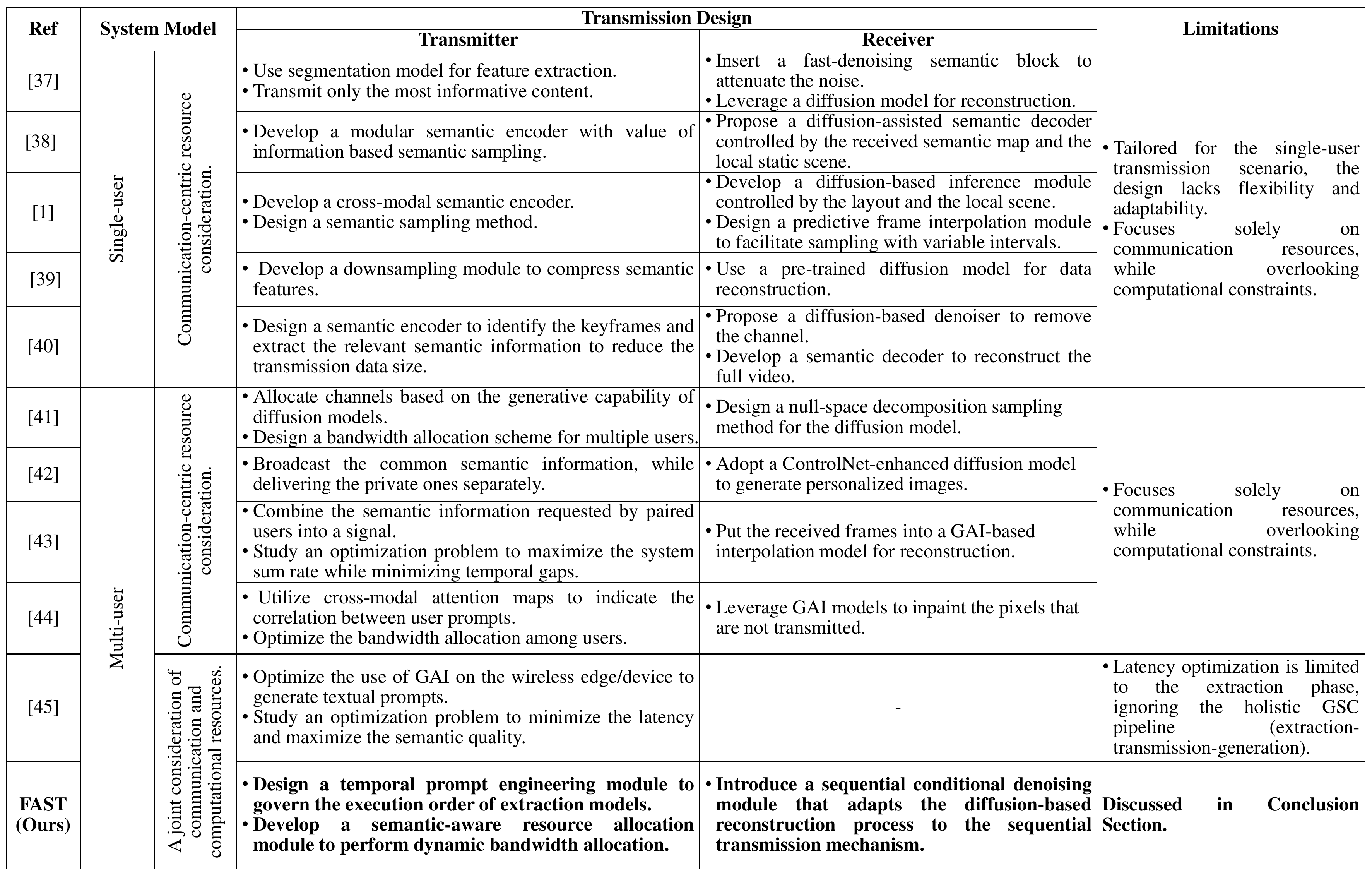}  \\
\end{tabular}
\label{table}
\vspace{-0cm}
\end{table*}

\section{System Model}\label{System Model}

\begin{figure*}
    \centering
    \includegraphics[width=18cm]{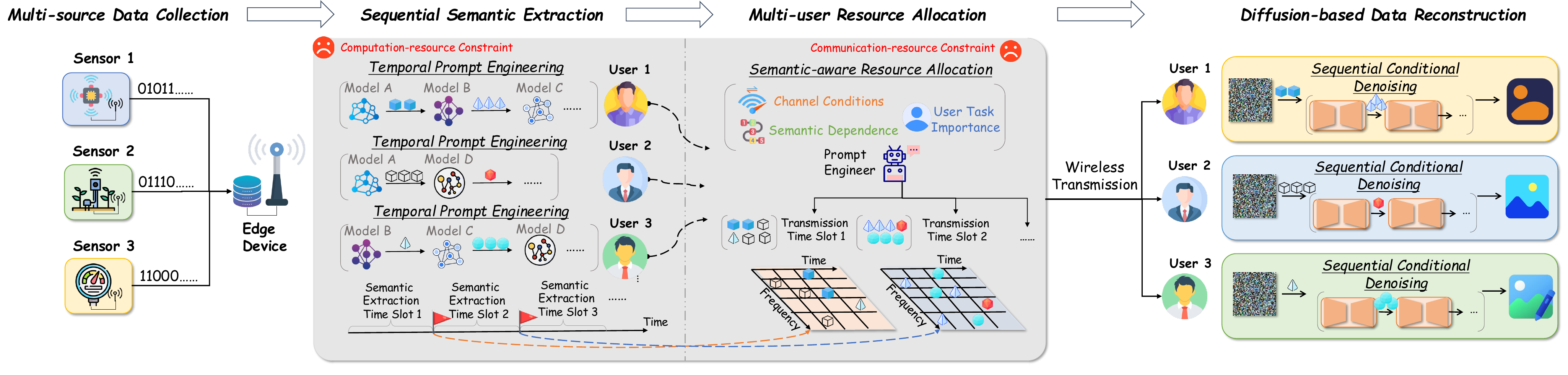}
    \caption{An overview of the proposed FAST-GSC framework.}
    \label{fig:workflow}
\end{figure*}

We consider a multi-user system where computational resources for semantic extraction and communication resources for wireless transmission are shared among all users. Image transmission is adopted as a representative use case, with semantic units defined as scene entities, their textual descriptions, and inter-entity relations. This formulation enables tractable experimentation while maintaining scalability to more complex scenarios. The workflow and key components of the proposed FAST framework are illustrated in Fig. \ref{fig:workflow}.

The proposed FAST framework operates through four sequential stages: 1) multi-source data collection, 2) sequential semantic extraction, 3) multi-user resource allocation, and 4) diffusion-based data reconstruction.

Initially, the edge server collects raw data from nearby sensors. Subsequently, semantic extraction models are executed in an order determined by the temporal prompt engineering module. This sequential approach is designed to efficiently utilize constrained computational resources at the edge. During the multi-user resource allocation stage, the allocated communication resources are dynamically optimized for the transmitted semantic units by the semantic-aware resource allocation module, factoring in semantic dependencies, the task importance of each user, and real-time channel conditions.

At the user end, a diffusion model initiates the reconstruction process immediately upon the arrival of the first semantic units, thereby reducing latency. This reconstruction is iteratively refined at each time step, guided by all semantic units received up to that point. As the units arrive sequentially, the guidance signal for the diffusion model evolves across its denoising steps. To accommodate this dynamic guidance, we introduce a sequential conditional denoising module, which enables the model to efficiently integrate the progressively arriving semantic information.

\section{Key Components of FAST}\label{components}

The temporal prompt engineering and sequential conditional denoising modules form a complementary pair: the former regulates the sequential transmission order at the transmitter, and the latter accommodates this sequence at the receiver. Designed independently yet coordinated for seamless operation, this per-user unit is elaborated first. Following this, we scale the framework to multi-user transmission, detailing the integrated semantic-aware resource allocation module for resource allocation.

\subsection{Temporal Prompt Engineering}\label{TPE}

The semantic extraction process is dedicated to deriving concise semantic representations from the original source data. To accommodate heterogeneous source data distributions and diverse user demands, multiple specialized semantic extraction models can be employed to distill multi-dimensional semantic units. Herein, each extraction result is denoted as a semantic unit. For a given user, the collection of all semantic units forms an ordered prompt sequence, which we inject progressively into the diffusion process over successive denoising steps.

We denote the source collected from sensor $i$-th as ${\mathbf{s}_i} \in {\mathbb{R}^{c \times h \times w}}$. The semantic extraction process of employing the $k$-th extraction model for distilling the semantic units ${{\mathbf{u}}_{ki}}$ from ${\mathbf{s}_i}$ can be expressed by
\begin{equation}
    {{\mathbf{u}}_{ki}} = {f_k}\left( {{\mathbf{s}_i};{{\bm{\psi }}_k}} \right),
\end{equation}
where ${{{\bm{\psi }}_k}}$ represents the trained parameters of the extraction model $k$, $K$ represents the maximum number of the semantic extraction models and $\forall k \in \left\{ {0,1,...,K - 1} \right\}$. 

However, the execution of a large number of extraction models imposes significant challenges in computation resource-constrained scenarios, potentially resulting in prolonged transmission latency and excessive computational memory consumption. Therefore, we propose to execute the semantic extraction models sequentially to distill key semantic units for transmission. 

Rather than collecting all semantic units for every user prior to transmission, the semantic units distilled at the current timestamp undergo a multi-user resource allocation procedure and are transmitted to their respective target users immediately. Meanwhile, the diffusion model’s denoising process is initiated by the first received semantic units. In subsequent timestamps, newly arrived semantic units combine with previously arrived units to collectively guide the denoising process.

\begin{figure*}
	\centering
	\includegraphics[width=15cm]{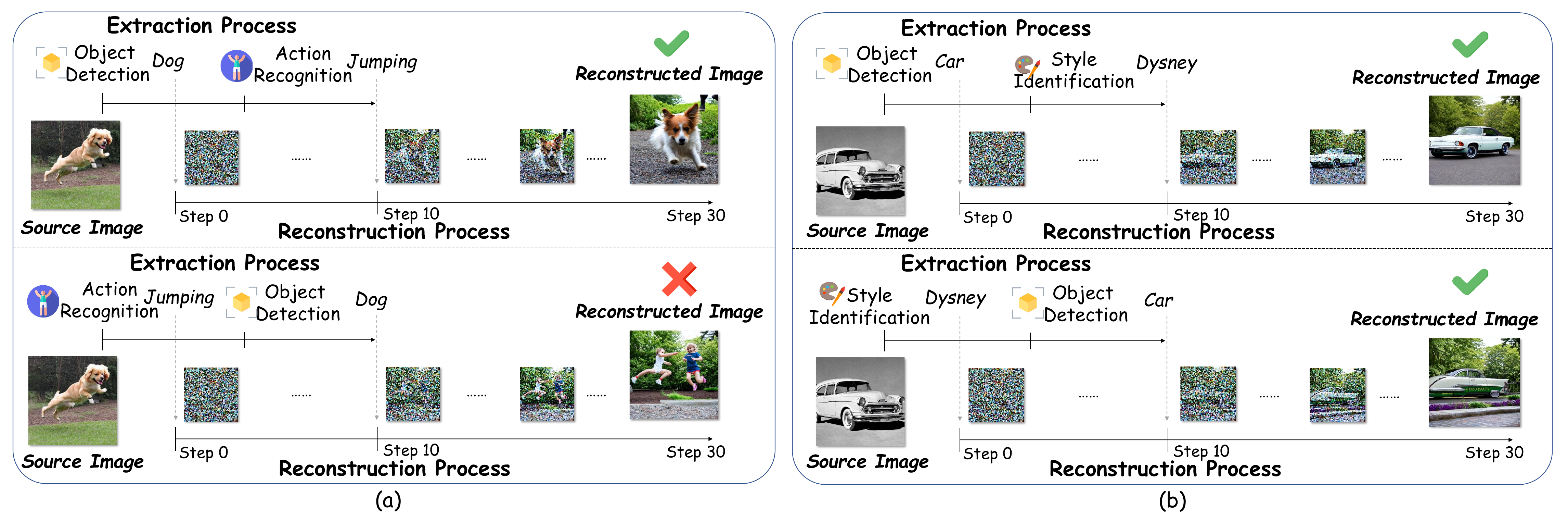}
	\caption{Illustrative examples of some direct implementations of the sequential transmission method. The total denoising step is set at 30 for all schemes. In (a), the switch of the execution order of these two extraction models yields semantically different results. In (b), the switch of the execution order of the two extraction models doesn't change the semantic meanings of reconstruction results.}
	\label{fig:fig2_1}
\end{figure*}

However, initial experimental results indicate the presence of sequential dependencies among semantic units, suggesting that the execution order of extraction models significantly affects the quality of the reconstructed data. As illustrated in Fig. \ref{fig:fig2_1} (a), with the same extraction speed, changing the extraction and transmission sequence between two semantic units can yield significantly different outcomes. This phenomenon demonstrates the tight sequential dependency between these two semantic units during the reconstruction process of the diffusion model. In contrast, as shown in Fig. \ref{fig:fig2_1} (b), switching the extraction and transmission sequence between another pair of semantic units results in similar reconstruction outcomes both semantically and visually, indicating a weak sequential dependency between these units.

Given the aforementioned observation that the order of extraction and transmission of different semantic units should follow their sequential dependencies, we resort to RL to learn and implement this strategy. Its aim is to ensure the final inferred images bear a close resemblance to the comprehensive prompt, while simultaneously minimizing system latency as much as possible. The designs of the fundamental components are outlined in the following content.

\begin{figure}
    \centering
    \includegraphics[width=9.1cm]{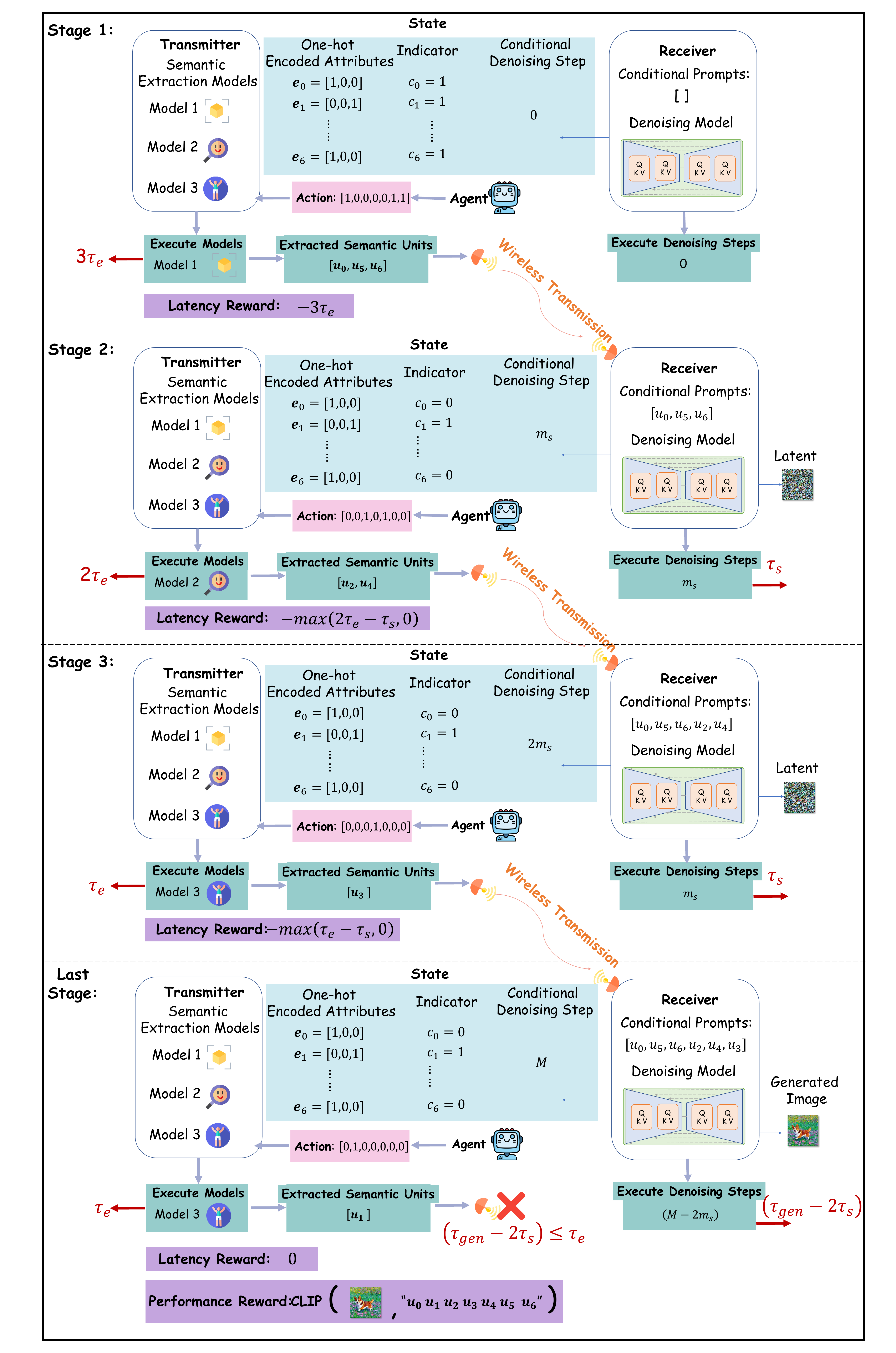}
    \caption{Illustration depicting an example of a single episode within the MDP framework. In the first RL phase, all indicators are set to one, and the agent selects the extraction model to be executed at this phase. As the denoising process has not yet started, these semantic units can be integrated into the diffusion model from the initial denoising step. At the beginning of the second RL phase, the extracted results distilled at the first RL phase arrive at the receiver and initiate the denoising process. The indicators corresponding to the transmitted units are set to zero. In the final phase, the denoising process reaches its scheduled total phase and terminates the episode. Given that the extraction latency surpasses the denoising duration at this phase, one semantic unit remains unsuccessfully transmitted.}
    \label{fig:fig3}
\end{figure}

\textit{1) MDP Constructing:} In constructing the RL model, we frame the interaction between the agent and the environment as a Markov Decision Process (MDP). The MDP can be defined as a three-component tuple $\left( {\mathcal{S},\mathcal{A},r} \right)$, where $\mathcal{S}$ is state space set, $\mathcal{A}$ is the action space set, and $r$ is the reward function. Within this framework, an episode is defined as a complete trajectory from the initial state to the terminal state, consisting of a sequence of states, actions, and corresponding rewards. To simplify the learning process of RL models, we divide the reconstruction process into multiple segments at equal intervals. At the beginning of each reconstruction segment, only newly arrived semantic units can be integrated with previously arrived ones to guide the denoising. This approach allows the RL model to learn reception deadlines of semantic units with different attributes, thereby establishing the extraction and transmission sequence. To better illustrate the MDP in our study, we depict an episode in Fig. \ref{fig:fig3} and the detailed design of this framework is given as follows:

\textbf{State:} We include the task-related request in the agent's state, which contains the attribute and quantity of semantic attributes required by the receiver. To facilitate the learning of the RL model, we can employ one-hot encoding to preprocess the semantic attributes. Let \(\mathbf{e}_{k,t}\) denote the encoded $N_\text{e}$-length one-hot vector for the $k$-th required semantic attribute at phase\footnote{In our work, the term `phase’ corresponds to a single iteration or transition within the RL environment, with each phase forming a segment of an episode.} $t$, where the $n_\text{e}$-th position is set to 1, corresponding to the activation of the $n_\text{e}$-th extraction model to obtain the relevant semantic unit. When the requested number of semantic units is less than $K$, we fill the remaining positions in the one-hot encoding with zero vectors. Subsequently, the entire encoded one-hot matrix for the whole prompt can be expressed by \(\mathbf{E}_t = \left[ \mathbf{e}_{0,t}, \ldots, \mathbf{e}_{K-1,t} \right]\). Since the residual noise levels vary at different denoising steps, and the allowable space for modifications also changes, the decision-making process of the RL model should include the conditional denoising step of the diffusion model \(d_t\) as part of its state, which indicates the starting denoising step from which the semantic units selected at time $t$ can guide. When \(d_t\) reaches the designated maximum number of denoising steps, it signifies the conclusion of the episode. To distinguish whether the semantic units are still required by the receiver at phase $t$, we also design a binary \(K\)-length vector ${{{\mathbf{c}}_t}}$ as an indicator. When an element of this vector is set to 0, it indicates that the corresponding semantic unit has either been transmitted or the associated one-hot attribute vector is a zero vector. The element is set to 1 otherwise. Therefore, the corresponding state observed by the agent at time phase \(t\) is denoted as ${{\mathbf{s}}_t} = \left[ {{{\mathbf{E}}_t},{{\mathbf{c}}_t},{d_t}} \right]$.

\textbf{Action:} The action executed by the agent at time $t$ is denoted by ${{\mathbf{a}}_t} = \left[ {{a_0},...,{a_{K - 1}}} \right]$, where ${a_k} \in \left\{ {0,1} \right\}$ for $k = 0,...,K - 1$. The practical implication of this action is that, when the agent selects certain semantic attributes, the corresponding extraction models that distill the semantic units with the selected attributes are executed at this RL phase. 
    
\textbf{Reward:} Aligned with our objectives of ensuring high reconstruction precision while reducing task latency, the reward is a composite of two components: the system's \textit{performance reward} and the \textit{latency reward}. Similar to \cite{harnessing}, we adopt the CLIP score \cite{clipscore} as the performance reward in our work. This metric assesses the semantic similarity between the inferred images and the corresponding prompts within the embedding space of CLIP model \cite{clipmodel}, which can be expressed by
\begin{equation}
    {r_\text{{p}}} = \operatorname{CLIP- S} = w_1 \cdot \max \left( {\cos \left( {{\mathbf{h}},{\mathbf{v}}} \right),0} \right),
\label{clip_score_function}
\end{equation}
where ${\mathbf{h}}$ represents the prompt's  embedding, ${\mathbf{v}}$ denotes visual 
embedding and $w_1 = 1$ in our work. Given that the intermediate results in the denoising process are inherently noisy \cite{DDPM, DDIM, stable}, they can not directly reflect the semantics and qualities of the final images. Consequently, the reward is calculated and assigned to the agent only after the completion of the entire denoising process\footnote{Should an efficient and rapid methodology for assessing semantics from partially denoised images be established, it would enable the deployment of some reward shaping techniques \cite{sparse1, sparse2} for addressing this sparse reward and accelerating the training of the RL model. This area of research remains open for future studies.}.

Building on this framework, system latency is determined by the degree of parallel execution between semantic extraction and reconstruction processes. Specifically, in the first phase of each episode, the receiver has not yet received any semantic units, and thus, denoising has not yet started. The latency of this phase is entirely determined by the execution time of the selected extraction model. From the second phase onward, the receiver can condition on the received semantic units, and extraction at the transmitter and reconstruction at the receiver start simultaneously. The latency of these phases is determined by the maximum latency of either the transmitter's semantic extraction or the receiver's segmented denoising. Thus, instead of calculating system latency at the conclusion of one episode, we can expedite the learning process by defining the latency reward as the additional extraction delay relative to the segmented denoising duration at each phase. Therefore, the combined reward obtained at phase $t$ can be formulated as
\begin{equation}
  {r_{t}} = \left\{ {\begin{array}{*{20}{c}}
  -{{\tau _\text{e}} \cdot {n_t}}&{t = 0,} \\ 
  -{\max \left( {\left( {{\tau _\text{e}} \cdot {n_t} - {\tau _\text{s}}} \right),0} \right)}&{1 \leqslant t < L - 1,} \\ 
  -{\max \left( {{{\mathbf{1}}_{nor}} \cdot \left( {{\tau _\text{e}} \cdot {n_t} - {\tau _\text{s}}} \right),0} \right) + {r_p}}&{t = L - 1,} 
\end{array}} \right.
\end{equation}
where ${\tau}_\text{e}$ denotes the latency for extracting a single semantic unit, ${n_t}$ denotes the number of selected semantic units at RL phase $t$, ${\tau}_\text{s}$ is time delay for executing the segmented denoising steps. $L$ denotes the number of phases in one episode, which varies with each episode. Its value depends on the number of semantic units required for transmission in each round and the progress of model learning. The index factor, ${{{\mathbf{1}}_{nor}}}$, equals 1 if the distilled semantic units at this phase can be received before the conclusion of the denoising process, and 0 otherwise. 

Finally, the long-term reward for the agent can be calculated as the sum of the discounted rewards obtained at each phase in an episode, which is expressed by
\begin{equation}
    r = \sum\limits_{t = 0}^{L - 1} {{\gamma ^t} \cdot {r_{t}}} ,
\end{equation}
where $\gamma $ is a discount factor.

\textit{2) Learning Principle:} Building on the MDP framework described above, we propose utilizing the Proximal Policy Optimization (PPO) \cite{ppo} algorithm to identify the optimal strategy for dispatching semantic units. PPO is a policy gradient method that directly updates a stochastic policy neural network,  ${\pi _{\boldsymbol{\varphi}} }$, to model the action distribution for a given state as ${\pi _{\boldsymbol{\varphi}} }\left( {\mathbf{s},\mathbf{a}} \right)$. Generally, the PPO algorithm employs three neural networks: the current policy network ${\pi _{\boldsymbol{\varphi}}}$ with parameter $\bm{\varphi} $, the last updated policy network ${\pi _{\boldsymbol{\varphi '}}}\left( {{\mathbf{s}},{\mathbf{a}}} \right)$ with parameter ${\boldsymbol{\varphi '}}$ and a critic network ${V_{\boldsymbol{\phi}} }$ with parameter ${\boldsymbol{\phi}}$. Both the new and old policy networks generate the action's probability distribution under the current state, with the old policy network serving to constrain the variance of the new policy. The critic network estimates the state value function for the current state, which assesses the performance of the new policy network.

To efficiently leverage experiences gathered under an old policy ${{\pi _{\boldsymbol{\varphi '}}}\left( {\left. {{{\mathbf{a}}_t}} \right|{{\mathbf{s}}_t}} \right)}$ for estimating potential performance under a new policy ${{\pi _{\boldsymbol{\varphi }}}\left( {\left. {{{\mathbf{a}}_t}} \right|{{\mathbf{s}}_t}} \right)}$, we can define the probability ratio between the new policy and the old policy as ${r_t}\left( {\boldsymbol{\varphi }}  \right) = \frac{{{\pi _{\boldsymbol{\varphi}} }\left( {\left. {{{\mathbf{a}}_t}} \right|{{\mathbf{s}}_t}} \right)}}{{{\pi _{\boldsymbol{\varphi '}}}\left( {\left. {{{\mathbf{a}}_t}} \right|{{\mathbf{s}}_t}} \right)}}$. Subsequently, the clipped surrogate objective in PPO algorithm is formulated by
\begin{equation}
    {\rm \mathcal{L}}_t^{\text{clip}}({\boldsymbol{\varphi}}) = \hat{\mathbb{E}}_t \left[ \min(r_t({\boldsymbol{\varphi }}) \hat{A}_t, \text{clip}(r_t({\boldsymbol{\varphi}}), 1 - \eta, 1 + \eta ) \hat{A}_t) \right],
\end{equation}
where $\text{clip}\left(  \cdot  \right)$ is the clip function which prevents the new policy from going far away from the old policy by limiting the probability ratio to the interval $\left[ {1 - \eta ,1 + \eta } \right]$. $ \hat{A}_t$ represents the estimated advantage function, given by $ \hat{A}_t = \delta_t + (\gamma \lambda) \delta_{t+1} + \ldots + (\gamma \lambda)^{T-t+1} \delta_{T-1}$, where $\gamma$ is the discount factor and $\lambda$ is the trace-decay parameter. $\delta_t = r_t + \gamma V_{\boldsymbol{\phi}}(\mathbf{s}_{t+1}) - V_{\boldsymbol{\phi}}(\mathbf{s}_t)$ represents the temporal difference error, and $V_{\boldsymbol{\phi}}(\mathbf{s}_t)$ is the state value function output by the critic network with parameter ${\boldsymbol{\phi}} $.  The critic network is updated by gradient descent, i.e. ${\boldsymbol{\phi}}  = {\boldsymbol{\phi}}  - \alpha {\nabla _{\boldsymbol{\phi}} }{\rm \mathcal{L}}\left( {\boldsymbol{\phi}}  \right)$, where ${\rm \mathcal{L}}\left( {\boldsymbol{\phi}}  \right)$ is the loss function for the critic network, which is a mean squared error function defined as ${\rm \mathcal{L}}({\boldsymbol{\phi}}) = \mathbb{E} \left[ \left( \sum_{\tau=0}^{\infty} \gamma^\tau r_{t+1+\tau} - V_{{\boldsymbol{\phi}}}(\mathbf{s}_t) \right)^2 \right]$.

Moreover, to encourage exploration and prevent the agent from falling into a suboptimal situation, we incorporate the entropy of the probability distribution into the loss function to increase the policy’s exploration of actions. Finally, the loss function for the new policy network can be expressed by
\begin{equation}
    {\rm \mathcal{L}}_t^{\text{PPO}}({\boldsymbol{\varphi}}) = {\rm \mathcal{L}}_t^{\text{clip}}({\boldsymbol{\varphi}}) + \xi \cdot {S_{{\pi _{\boldsymbol{\varphi}} }}}\left( {{{\mathbf{s}}_t}} \right),
\end{equation}
where ${S_{{\pi _{\boldsymbol{\varphi}} }}}\left( {{{\mathbf{s}}_t}} \right)$ is the entropy of the new policy network at time phase $t$ and $\xi$ denotes the entropy coefficient.


\textit{3) Invalid Action Masking:}
In complex task scenarios, the receiver may require a substantial number of semantic units. Our action space, which expands exponentially with the maximum required number of semantic units for transmission, can significantly impede the learning process. To streamline this, we propose excluding certain actions at each RL phase that fail to provide valuable semantic information to the receiver. These actions do not contribute to the integration of new semantic attributes into the inferred images, thereby not enhancing the performance score. Furthermore, these actions also squander transmission resources that could be more effectively allocated to conveying other informative semantic units within the specified transmission deadlines. In our work, we identify two types of non-contributory actions: 1) Actions intended to distill semantic units associated with a zero-vector in their one-hot attribute representation; and 2) Actions aimed at distilling semantic units that have already been extracted and transmitted. By removing these non-contributory actions from the action space, we can simplify and accelerate the learning process of the RL model. The effectiveness of this approach is validated through simulations, as detailed in Section \ref{TPE_SIM}.

To achieve this, we utilize the invalid action masking technique \cite{mask}, which is commonly implemented to avoid repeatedly generating invalid actions in large discrete action spaces. Specifically, we set the logits associated with the aforementioned non-contributory actions to $ - \infty $ and normalize the probabilities of all the actions by
\begin{equation}
    {\text{softmax}}\left( {{z_i}} \right) = \frac{{{e^{{z_i}}}}}{{\sum\limits_{j = 0}^{{N_{\text{A}}}} {{e^{{z_j}}}} }},
\end{equation}
where ${{z_i}}$ is the model’s output probability distribution of action ${{\mathbf{a}}_i}$.  ${{N_{\text{A}}}}$ denotes the total number of actions in the action set, which equals to ${2^K}$ in our work. The probability of sampling invalid actions is equal to 0 as $\mathop {\lim }\limits_{z \to  - \infty } {e^z} = 0$.

\subsection{Sequential Conditional Denoising}
Inspired by the significant success of diffusion models across a wide range of real-world generative reconstruction tasks, the Stable Diffusion model \cite{stable} is commonly employed for generative reconstruction at the receiver within the GSC framework \cite{harnessing, xia}. Its goal is to generate images that semantically adhere to the provided prompt.

The stable diffusion model is built upon the denoising diffusion probabilistic model (DDPM) \cite{DDPM}, which is capable of learning a distribution ${p_\theta }\left( {{{\mathbf{x}}_0}} \right)$ that approximates $q\left( {{{\mathbf{x}}_0}} \right)$. It can be treated as a latent variable model of the form ${p_\theta }\left( {{{\mathbf{x}}_0}} \right) = \int {{p_\theta }\left( {{{\mathbf{x}}_{0:T}}} \right)} d{{\mathbf{x}}_{1:T}}$, where ${{\mathbf{x}}_1},...,{{\mathbf{x}}_T}$ are called latents of the same dimensionality as the original data ${{\mathbf{x}}_0} \sim q\left( {{{\mathbf{x}}_0}} \right)$. The joint distribution ${p_\theta }\left( {{{\mathbf{x}}_{0:T}}} \right)$ is called the \textit{reverse process}, which can be formulated as ${p_\theta }\left( {{{\mathbf{x}}_{0:T}}} \right) = p\left( {{{\mathbf{x}}_T}} \right)\mathop \prod \limits_{t = 1}^T {p_\theta }\left( {\left. {{{\mathbf{x}}_{t - 1}}} \right|{{\mathbf{x}}_t}} \right)$. 

The learning process starts from a pure noise image ${{\mathbf{x}}_T} \sim {\rm \mathcal{N}}\left( {{{\mathbf{x}}_T};{\mathbf{0}},{\mathbf{I}}} \right)$. The Markov chain with learned Gaussian transition can be expressed by ${p_\theta }\left( {\left. {{{\mathbf{x}}_{t - 1}}} \right|{{\mathbf{x}}_t}} \right) = {\rm \mathcal{N}}\left( {{{\mathbf{x}}_{t - 1}};{{\mathbf{\mu }}_\theta }\left( {{{\mathbf{x}}_t},t} \right),{{\mathbf{\Sigma }}_\theta }\left( {{{\mathbf{x}}_t},t} \right)} \right)$. To facilitate the learning process, a \textit{forward process} or \textit{diffusion process} $q\left( {\left. {{{\mathbf{x}}_{1:T}}} \right|{{\mathbf{x}}_0}} \right) = \mathop \prod \limits_{t = 1}^T q\left( {\left. {{{\mathbf{x}}_t}} \right|{{\mathbf{x}}_{t - 1}}} \right)$ is implemented, which is a fixed Markov chain that gradually adds Gaussian noise to the data according to a variance schedule ${\beta _1},...,{\beta _T}$, as $q\left( {\left. {{{\mathbf{x}}_t}} \right|{{\mathbf{x}}_{t - 1}}} \right) = {\rm \mathcal{N}}\left( {{{\mathbf{x}}_t};\sqrt {1 - {\beta _t}} {{\mathbf{x}}_{t - 1}},{\beta _t}{\mathbf{I}}} \right)$. Denote ${\alpha _t} = 1 - {\beta _t}$ and ${{\bar \alpha }_t} = \mathop \prod \limits_{s = 1}^t {\alpha _s}$, we can sample ${{{\mathbf{x}}_t}}$ at timestep $t$ as $q\left( {\left. {{{\mathbf{x}}_t}} \right|{{\mathbf{x}}_0}} \right) = {\rm \mathcal{N}}\left( {{{\mathbf{x}}_t};\sqrt {{{\bar \alpha }_t}} {{\mathbf{x}}_0},\left( {1 - {{\bar \alpha }_t}} \right){\mathbf{I}}} \right)$. Subsequently, the forward posteriors conditioned on ${{{\mathbf{x}}_0}}$ can be expressed by
\begin{equation}
   q\left( {\left. {{{\mathbf{x}}_{t - 1}}} \right|{{\mathbf{x}}_t},{{\mathbf{x}}_0}} \right) = {\rm \mathcal{N}}\left( {{{\mathbf{x}}_{t - 1}};{{{\mathbf{\tilde \mu }}}_t}\left( {{{\mathbf{x}}_t},{{\mathbf{x}}_0}} \right),{{\tilde \beta }_t}{\mathbf{I}}} \right),
\end{equation}
where ${{{\mathbf{\tilde \mu }}}_t}\left( {{{\mathbf{x}}_t},{{\mathbf{x}}_0}} \right) = \frac{{\sqrt {{{\bar \alpha }_{t - 1}}} {\beta _t}}}{{1 - {{\bar \alpha }_t}}}{{\mathbf{x}}_0} + \frac{{\sqrt {{{\bar \alpha }_t}} \left( {1 - {{\bar \alpha }_{t - 1}}} \right)}}{{1 - {{\bar \alpha }_t}}}{{\mathbf{x}}_t}$ and ${{\tilde \beta }_t} = \frac{{1 - {{\bar \alpha }_{t - 1}}}}{{1 - {{\bar \alpha }_t}}}{\beta _t}$. The target of the learning process of DDPM is to minimize the gap between $q\left( {\left. {{{\mathbf{x}}_{t - 1}}} \right|{{\mathbf{x}}_t},{{\mathbf{x}}_0}} \right)$ and ${p_\theta }\left( {\left. {{{\mathbf{x}}_{t - 1}}} \right|{{\mathbf{x}}_t}} \right)$. When the coefficient of variance is considered as a constant, this goal is equivalent to predicting ${{{\mathbf{\mu }}_\theta }\left( {{{\mathbf{x}}_t},t} \right)}$ that as accurately as possible to match ${{{{\mathbf{\tilde \mu }}}_t}\left( {{{\mathbf{x}}_t},{{\mathbf{x}}_0}} \right)}$. Denote ${\mathbf{y}}$ as the prompt, the loss function during training is formulated as the difference between the generated noise $\varepsilon$ and predicted noise ${\varepsilon _\theta }\left( {{{\mathbf{x}}_0},t,{\mathbf{y}}} \right)$, as follows \cite{stable}:
\begin{equation}
    {\rm \mathcal{L}} = {{\rm {\mathbb{E}}}_{t,{\mathbf{y}},{{\mathbf{x}}_t},\varepsilon }}\left[ {\left\| {\varepsilon  - {\varepsilon _\theta }\left( {\sqrt {{{\bar \alpha }_t}} {{\mathbf{x}}_0} + \sqrt {1 - {{\bar \alpha }_t}} \varepsilon ,t, {\mathbf{y}} } \right)} \right\|_2^2} \right].
\end{equation}

Conventionally, during the sampling process, noise estimation is typically performed using classifier-free guidance \cite{classifier}, which can be expressed as
\begin{equation}
    {{\tilde \varepsilon }_\theta }\left( {{{\mathbf{x}}_t},{\mathbf{y}},t} \right) = {\varepsilon _\theta }\left( {{{\mathbf{x}}_t},{\mathbf{y}},t} \right) + w \cdot \left( {{\varepsilon _\theta }\left( {{{\mathbf{x}}_t},{\mathbf{y}},t} \right) - {\varepsilon _\theta }\left( {{{\mathbf{x}}_t},t} \right)} \right), \label{guidance}
\end{equation} 
where $w$ is the guidance scale. 

\begin{figure}
	\centering
	\includegraphics[width=9cm]{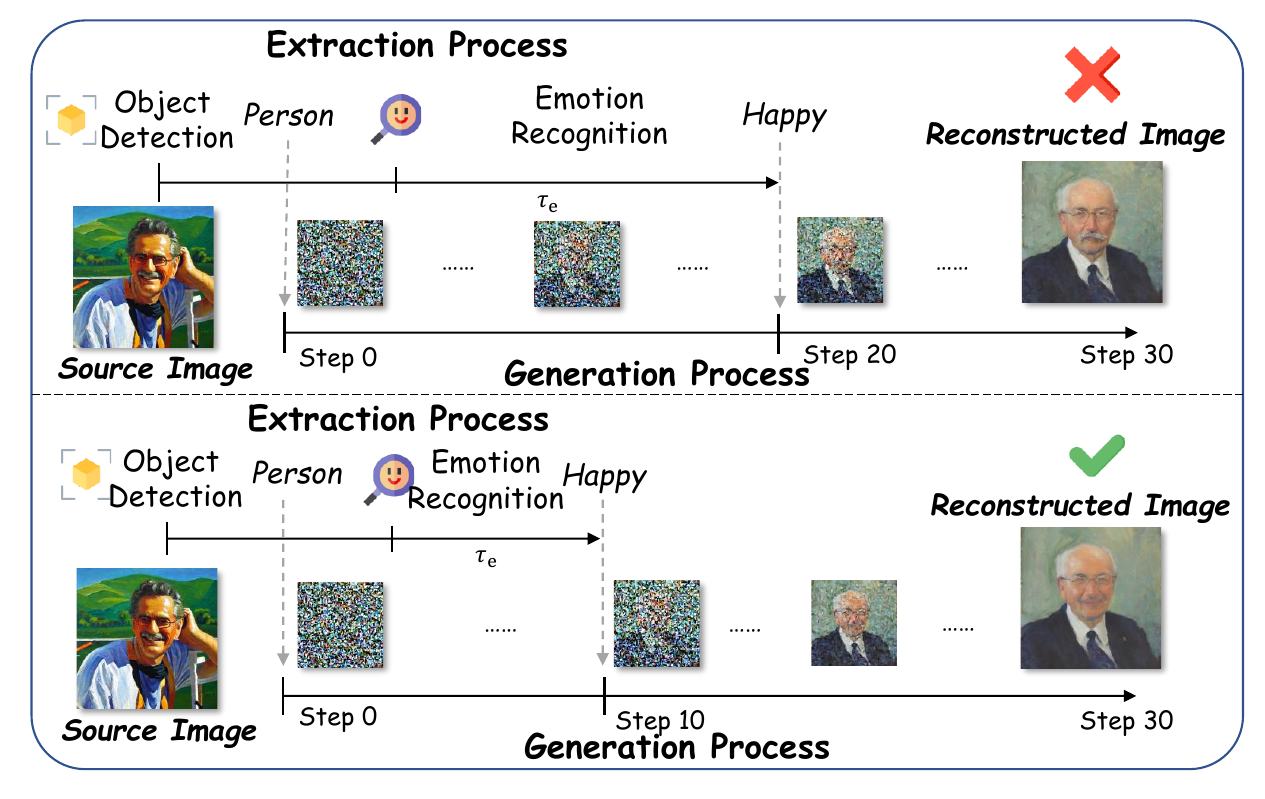}
	\caption{Illustrative examples of some direct implementations of the traditional denoising method. The total denoising step is set at 30 for all schemes. We posit that the speed of the emotion recognition model on the upper side is slower than on the lower side, leading to the emotion-related semantic unit on the upper side arriving 10 steps later than on the lower side. Observations indicate that the inferred image on the lower side effectively incorporates guidance from both extracted semantic units, whereas the upper one fails to integrate the second semantic feature.}
	\label{fig:fig2_2}
\end{figure}


In FAST, the denoising process begins after the reception of the first semantic unit. As the denoising process progresses, the noise level diminishes, limiting the capacity to incorporate new semantic features. We conducted experiments to evaluate whether the traditional denoising method mentioned above is compatible with our transmission schemes discussed in Section \ref{TPE}. As depicted in the upper part of Fig. \ref{fig:fig2_2}, due to the excessive processing delays of the second extraction model, the receiver can only integrate the second semantic unit into the conditional guidance at the 20-th denoising step. The minimal remaining noise at the 20-th denoising step limits the modification capacity of the second semantic unit, resulting in a certain degree of semantic mismatch between the source and inferred images. If the reception timestamp of the second semantic unit is 10 denoising steps earlier, its influence can still be observed, as shown in the lower part of Fig. \ref{fig:fig2_2}. The results indicate that the reception immediacy of semantic units significantly impacts the quality of the reconstructed data.

\begin{figure*}
	\centering
	\includegraphics[width=16cm]{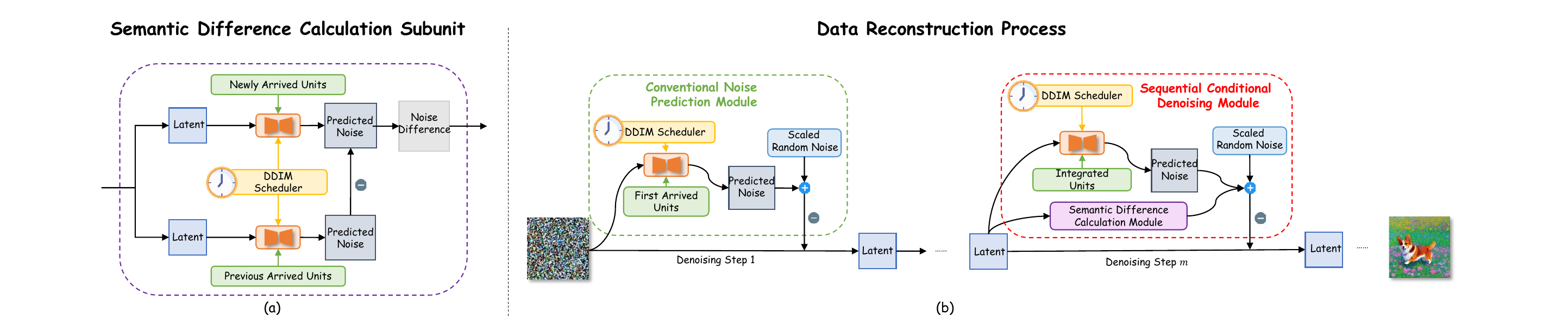}
	\caption{Demonstration of the data reconstruction process. (a) Workflow of the proposed semantic difference calculation subunit; (b) Illustration of the sequential conditional denoising module, where the guidance effectiveness of late-arrived semantic units is enhanced for improved final performance.}
	\label{fig: fig4}
\end{figure*}

Therefore, we aim to mitigate the stringent reception immediacy at each user end, thereby enabling the sequentially arrived units to continuously contribute to the reconstruction results. To achieve this goal, one straightforward method is to add a certain amount of noise to the reconstruction process every time a new semantic unit arrives at the receiver, ensuring enough space for modifications. However, this approach increases the overall noise level, requiring additional denoising steps and significantly prolonging the reconstruction process. 


In contrast, we devise a sequential conditional denoising module to enhance the intervention capability of late-arrived semantic units for final performance without disrupting the ongoing denoising schedule. Specifically, we first introduce a semantic difference calculation subunit that can output the difference between the estimated conditional noises guided by the newly received and the previously received semantic features, as depicted in Fig. \ref{fig: fig4} (a). Contrary to blindly introducing random noise to all the intermediate results, this subunit allows us to preciously and efficiently identify the space that the newly arrived semantic units target to modify at the current denoising step. Subsequently, we supplement the noise difference to the estimated noise conditioned on all the received semantic units using a weighted sum method. Building upon classifier-free guidance \cite{classifier} method, this modification can be formulated as follows:
\begin{align}
&{{\hat \varepsilon }_\theta }\left( {{{\mathbf{x}}_t},{{\mathbf{y}}_{{\text{combined}}}},{{\mathbf{y}}_{{\text{new}}}},{{\mathbf{y}}_{{\text{previous}}}},t} \right) \nonumber \\
& = {{\tilde \varepsilon }_\theta }\left( {{{\mathbf{x}}_t},{{\mathbf{y}}_{{\text{combined}}}},t} \right) \nonumber \\
& + \alpha  \cdot \left( {{\varepsilon _\theta }\left( {{{\mathbf{x}}_t},{{\mathbf{y}}_{{\text{new}}}},t} \right) - {\varepsilon _\theta }\left( {{{\mathbf{x}}_t},{{\mathbf{y}}_{{\text{previous}}}},t} \right)} \right), \label{weighting_equation}
\end{align}
where $\alpha$ is the intervention factor, ${{{\mathbf{y}}_{{\text{combined}}}}}$ denotes all the semantic units received at diffusion step $t$, ${{{\mathbf{y}}_{{\text{new}}}}}$ denotes the semantic units newly received at diffusion step $t$ and ${{{\mathbf{y}}_{{\text{previous}}}}}$ denotes the semantic units received before diffusion step $t$. 

Through this approach, the guidance effectiveness of late-arrived semantic units for final performance can be enhanced, which can mitigate the issue of many semantic features not being effectively incorporated into the reconstruction results due to their late extraction and transmission in complex tasks. The sequential conditional denoising module and the overall data reconstruction process are illustrated in Fig.~\ref{fig: fig4} (b). 


\subsection{Semantic-aware Resource Allocation}\label{SRA}

In conventional multi-user transmission systems, resource allocation is predominantly based on channel state information for bandwidth management. While effective for maximizing physical-layer metrics, they are insufficient for multi-user semantic transmission. This is because they fail to incorporate two critical dimensions of the problem: the complex semantic dependencies among information units, which dictate the value and decoding order of data, and the inherent disparity in task importance across users, which demands differentiated transmission priorities. Therefore, a novel strategy that jointly optimizes resources based on channel quality, semantic dependencies, and task priorities is essential.

Herein, we design an RL-based semantic-aware resource allocation module for dispatching the extracted semantic units. The key components are defined as follows:

\textbf{State:} For each user, we first compute the volume of the remaining extracted semantic units in the current time slot, categorized by semantic attribute, and store these values in a vector. We then append the user's signal-to-noise ratio to the vector, followed by an indicator reflecting the user’s task importance. The state representation is formed by concatenating the vectors from all users. Assuming there are $K$ distinctive semantic attributes and $M$ users, the resulting state vector has a dimensionality of $M * (K + 2)$.

\textbf{Action:} The action executed by the agent at time $t$ is denoted as ${{\mathbf{a}}_t} = \left[ {{a_0},...,{a_{M - 1}}} \right]$, where ${a_m}$ stands for normalized bandwidth allocation and ${a_m} \in \left[ {0,1} \right]$ for $m = 0,...,M - 1$.

\textbf{Reward:} Aligned with the object of ensuring higher task performance for users with greater task importance, the initial reward is defined as the weighted sum of all users' task performance:
\begin{equation}
    {r_s} = \frac{1}{M} \sum\limits_{m = 0}^{M - 1} {{w_{m}}} * {{r_{p,m}}}, 
\end{equation}
where ${{r_{p,m}}}$ denotes the performance score of user $m$, and ${{w_{m}}}$ represents the task importance of user-$m$. Same as the performance reward defined in the temporal prompt engineering module, ${{r_{p,m}}}$ can only be obtained at the end of the episode. 

Informed by the temporal prompt engineering module, we devise a reward shaping strategy by measuring the semantic dependency of each semantic unit. Semantic units that have a higher probability of being extracted first are defined as possessing the highest degree of semantic dependence. The intermediate reward is defined as the weighted sum of transmitted semantic dependence at this step, which is shown as follows:
\begin{equation}
    {r_d} = \frac{1}{M} \sum\limits_{m = 0}^{M - 1} {\sum\limits_{k = 0}^{K - 1} {{{w_{m}}} * {n_{m,k}}*{d_{m,k}}} },
\end{equation}
where ${{n_{m,k}}}$ is the number of user-$m$'s transmitted units of attribute $k$ and ${{d_{m,k}}}$ is the semantic dependence of attribute $k$.

Therefore, the combined reward obtained at step $t$ can be formulated as
\begin{equation}
    {r_t} = \left\{ {\begin{array}{*{20}{c}}
{{r_d},{\rm{ }}0 \le t < L - 1},\\
{{r_d} + {r_s},{\rm{ }}t = L - 1},
\end{array}} \right.
\end{equation}
where $L$ denotes the number of steps in one episode.

Building upon the MDP framework outlined above, we employ the PPO method described in Section \ref{TPE} to determine the optimized resource allocation strategy. A detailed discussion of this approach is omitted here for brevity.

\section{Performance Evaluation}\label{Experiments}

In this section, we implement the proposed FAST framework. The details of implementations and the analysis of the experimental results are also described.

\textbf{Datasets:} We utilize the text sentences from the DIFFUSIONDB \cite{diffusiondb} dataset as extracted prompts for both training and testing purposes. This dataset stands as the first extensive text-to-image prompt repository, exhibiting a wide array of syntactic and semantic attributes. We limit the length of the prompts to between 6 and 12 words, which reduces the dataset to approximately 0.2 million usable prompts. This dataset is divided into training and testing sets following a 9:1 ratio. We categorize each text word's semantic attribute into five types: NOUN, ADJECTIVE, STYLE\footnote{In our work, the style-related text words include 51 types of styles as detailed in \cite{design}, along with words that have proper noun properties.}, VERB, and OTHERS. Subsequently, we employ spaCy \cite{spaCy} to segment the prompts into semantic units and tag each unit correspondingly \cite{liu2024cross}.

\textbf{Experimental Settings:} We employ Stable Diffusion v1-4\footnote{https://huggingface.co/CompVis/stable-diffusion-v1-4} as the generative reconstruction model at the receiver and base our optimization strategies on it. The experiments are conducted on a server with an NVIDIA RTX 4090 GPU with 24 GB of memory. The operating system is Ubuntu 20.04 with Pytorch 2.0.1. We considered an extreme scenario where the computational resources at the edge are severely constrained, allowing for the execution of merely a single semantic extraction model per user at any given time. In the multi-user transmission setting, 5 users are uniformly distributed within a 10-meter radius of an edge server. The task importance scores assigned to the users are 5, 3, 1, 1, and 1, respectively. The large-scale fading is modeled as $- (35.3 + 37.6 * log10(d))$ in dB \cite{large_scale}, where $d$ is the distance between a user and the edge server in meters. For small-scale fading, Rayleigh fading model is adopted. The transmitting power and noise power are specified as $10$ dBm and $-80$ dBm, respectively. The maximum tolerable communication delay is set to 0.01 seconds. We assume the packet size is 256 bits for every semantic unit. The actual transmission rate is determined by the selected modulation and coding scheme \cite{Modulation}.

\textbf{Training Hyperparameters:}  We set the total number of denoising steps $M$ at 60. By default, we segment the entire process into intervals of 10 denoising steps and equate the extraction latency of distilling one text unit from the source image to 5 denoising steps. For the training of temporal prompt engineering and semantic-aware resource allocation modules, the learning rates are set to 0.009 and 0.00015, with entropy coefficients $\xi$ of 0.01 and 0.03, respectively. A shared discount factor $\gamma$ of 0.99 and a clip parameter $\eta$ of 0.2 are used for both.

\begin{figure}
	\centering
	\includegraphics[width=8cm]{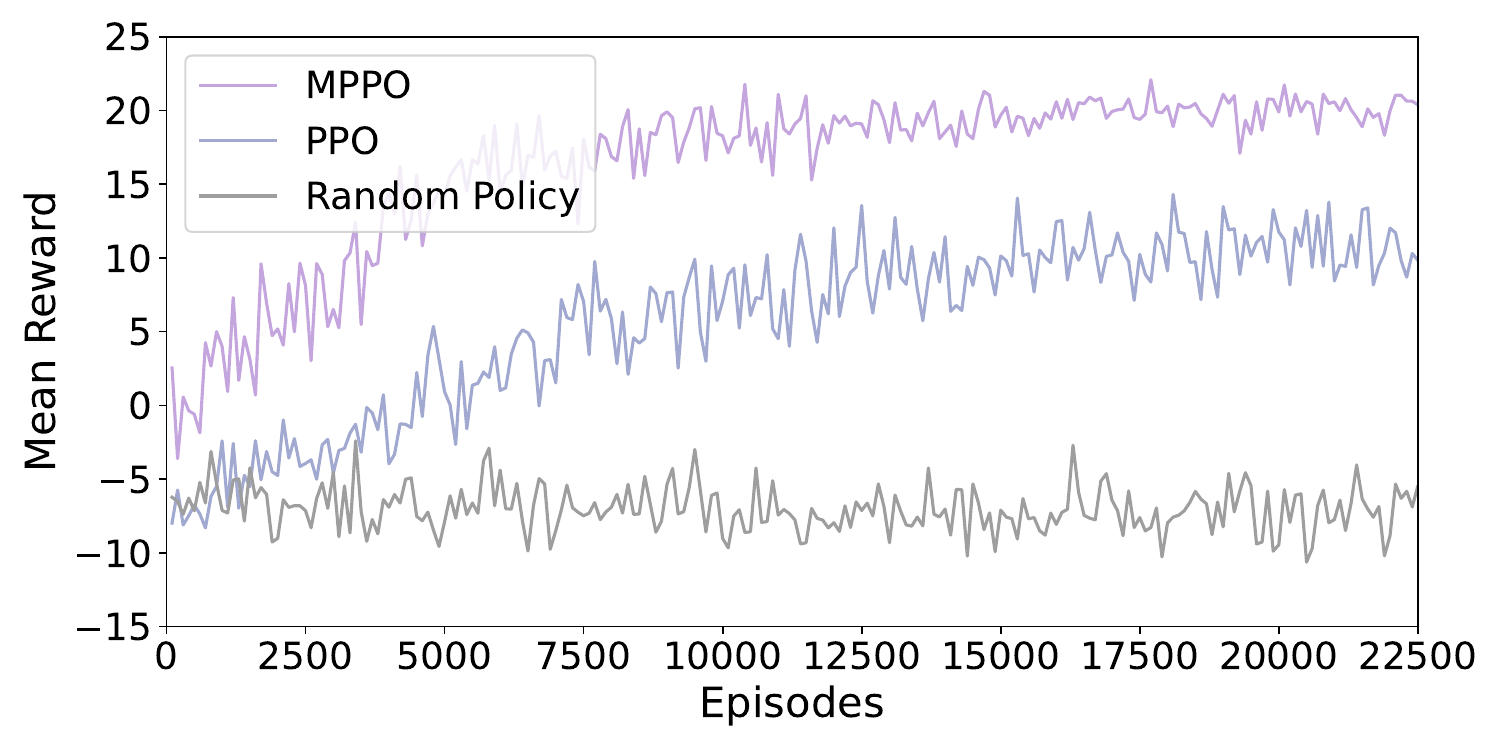}
	\caption{Learning performance of MPPO, PPO and random policy.}
	\label{fig:Convergence}
\end{figure}

\subsection{Evaluation of the Temporal Prompt Engineering Module}\label{TPE_SIM}
In this part, we evaluate the effectiveness of the proposed temporal prompt engineering module. Initially, we demonstrate the training process of our invalid action masking-assisted PPO model and compare its performance with that of two traditional methods. Subsequently, we conduct the statistical analysis of the trained model's results and investigate the presence of semantic patterns in its decision-making process.

\textit{1) Training Process:} The learning curves of the agents under different settings are shown in Fig. \ref{fig:Convergence}. We compare our invalid action masking-assisted PPO method (MPPO) against the conventional PPO method without masking and the random policy. It can be observed that our proposed MPPO algorithm offers notable enhancements compared to the one without masking in both convergence speed and the final values achieved. 
This superior performance can be attributed to the use of invalid action masking, which assists in eliminating non-contributory actions. This approach ensures that different semantic units are continually incorporated throughout the limited denoising process, enriching the content of the final inferred image and enhancing image quality. Additionally, as the number of transmissions increases, an increasing number of actions are identified and masked as non-contributory, leading to a faster learning speed and improved convergence of the algorithm.

\begin{figure*}
	\centering
	\includegraphics[width=16cm]{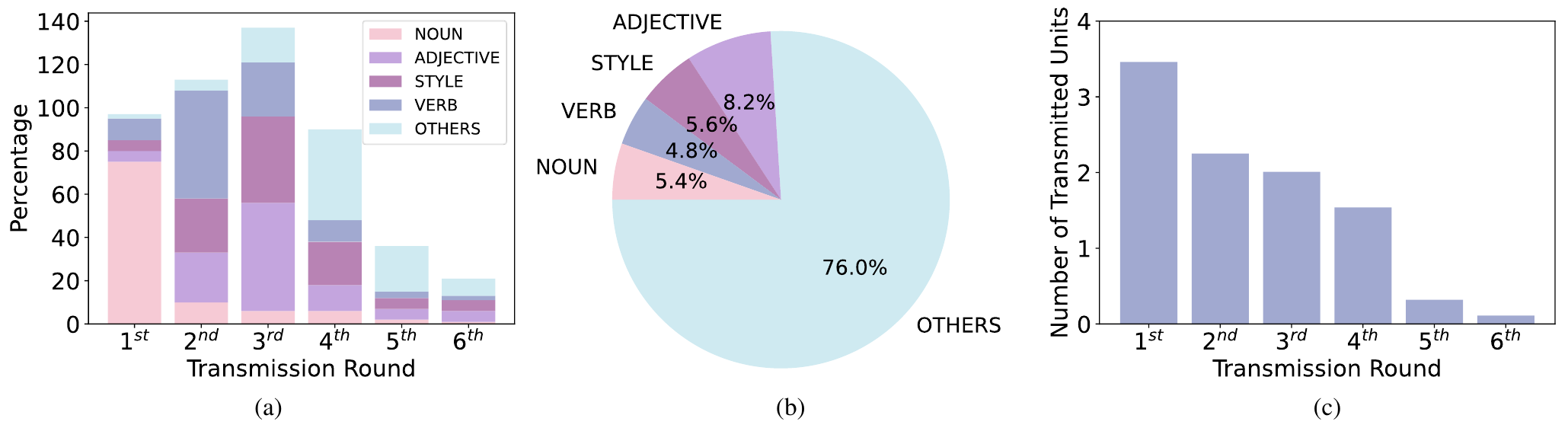}
	\caption{Statistical results of the temporal prompt engineering method. (a) Semantic characteristics of text units transmitted in chronological order; (b) Composition of discarded text units; (c) Number of text units transmitted in chronological order. According to our statistical results, six represents the maximum transmission round, as the number of text units transmitted from the seventh transmission onward is zero.}
	\label{fig:guideline}
\end{figure*}

\textit{2) Statistic Analysis:} Subsequently, we analyze the statistical results of the temporal prompt engineering module. First, we explore the semantic characteristics of text units sent in chronological order, as illustrated in Fig. \ref{fig:guideline} (a)\footnote{Note that to eliminate the impact of different proportions of text unit types in the dataset on the results, we compute the statistics based on their respective proportions over the time series. Consequently, the sum of their proportions at any given moment may exceed 100\%.}. These results reveal that the temporal prompt engineering module preferentially initiates the transmission of nouns. Consequently, extraction models that target noun retrieval, such as those used for object and scene recognition, should be prioritized in execution. During the second round of transmission, action-related units are more likely to be selected, corresponding to the outputs from action detection models. During the third round of transmission, units related to style and adjectives, which share similar characteristics, are predominantly selected. This indicates that the execution of perceptual extraction models, such as those used for sentiment analysis and style detection, can be appropriately delayed. Units of other types are typically chosen for transmission at later rounds due to their lesser contribution to the semantic information of the inferred images. 
Additionally, the results indicate that the obtained RL-based temporal prompt engineering module terminates the extraction process early and discards the transmission of a certain number of text units with a probability of 7.3\%. We present the components of the discarded text units in Fig. \ref{fig:guideline} (b), which shows that the majority of discarded units fall into the `OTHERS' category, such as `of', `a', and `the'. This indicates that in the current reconstruction task, these units represent semantic information that is less relevant to the task. Finally, we analyze the number of text units transmitted at different time intervals adopting the temporal prompt engineering module, as presented in Fig. \ref{fig:guideline} (c). The results indicate that a large amount of text units are transmitted during the earlier transmission. This is primarily because the semantic guidance effect is most effective at the initial stage of denoising. Consequently, the system prioritizes maintaining a high-performance score, even at the expense of some latency. In the subsequent transmission rounds, the number of transmitted text units decreases to approximately two. During these phases, the extraction latency and segmented denoising duration are closely aligned, indicating the effective parallelization of these processes. In the final transmission rounds, the number of transmitted text units significantly decreases. This reduction is primarily attributed to the smaller number of long prompts in the dataset and the discarding phenomenon of the temporal prompt engineering module.

\subsection{Evaluation of the Sequential Conditional Denoising Module}
In this section, we evaluate the performance of the sequential conditional denoising module. Initially, we present visual demonstrations of the inputs and outputs from the proposed semantic difference calculation subunit, showcasing how this subunit precisely amplifies the modifications required for newly received text units. Subsequently, we provide a visual analysis of the impact of the intervention factor $\alpha$ on the method's performance.

\begin{figure*}
	\centering
	\includegraphics[width=16cm]{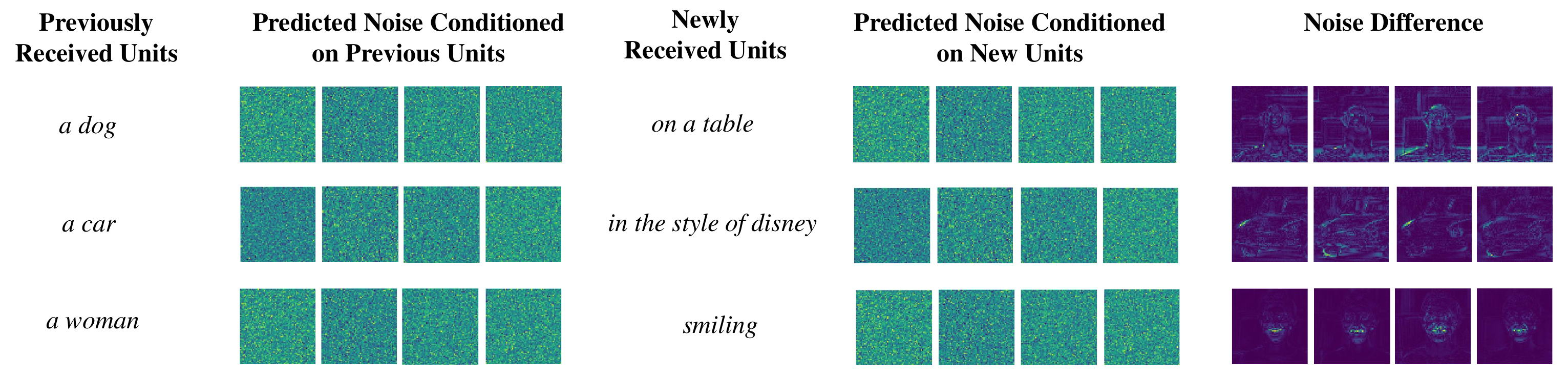}
	\caption{Detailed visual demonstration of the semantic difference calculation module's working process. For each row, the intermediate output of a representative image generation is presented. The visualization is achieved by unfolding the latent representation of the conditional noise predictions across the channel dimension (4 channels), revealing the distinct information embedded in each $64*64$ feature map. }
	\label{fig:noise}
\end{figure*}
\textit{1) Visual Demonstration of the Working Process of Semantic Difference Calculation Module:}
A visual representation of the working processes of the semantic difference calculation module is depicted in Fig. \ref{fig:noise}. We segment each prompt into two sets, displayed in the first and second columns, respectively. Initially, we transmit the units from the first column to the receiver to initiate the image reconstruction process. Subsequently, after a 20-step denoising period, we transmit the units from the second column. As shown in Fig. \ref{fig:noise}, direct observation of the predicted noise conditioned on both the previous and the new units may not provide useful information; however, the differences between them reveal significant semantic meanings. This distinction presents the necessary adjustments in the noise space to effectively incorporate the influence of late-arrived units on the reconstruction results. Specifically, in the first transmission scenario, depicted in the first row, the noise difference highlights the edge of the table and the dog's foot. In the second transmission scenario, the edge of the car, particularly the headlights and tires, is emphasized. This emphasis is attributed to the distinctive, rounded contours characteristic of objects depicted in the Disney style. In the third transmission scenario, the bright areas in the noise difference image represent the woman's mouth, which is the focal point for achieving the `smiling' directive. By employing this module, we can precisely identify the portions of the noise space that the newly arrived semantic features aim to modify.

\begin{figure*}
	\centering
	\includegraphics[width=16cm]{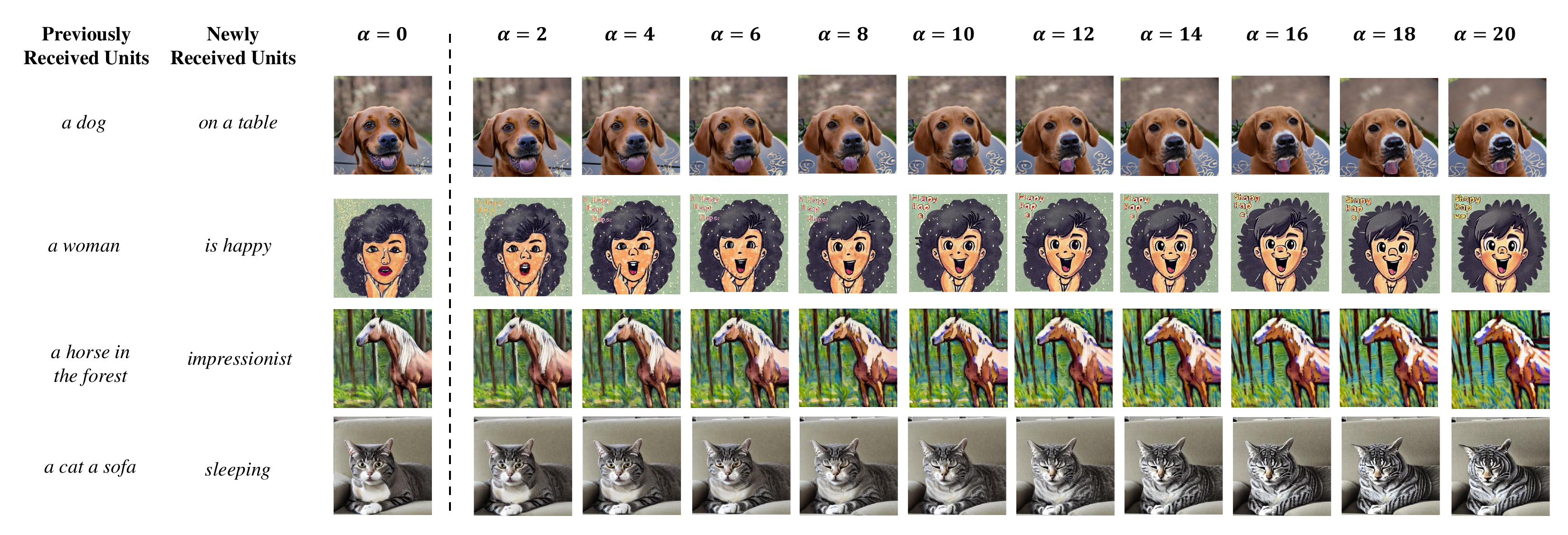}
	\caption{Visual illustrations of changes in intervention factor values on the inferred images.}
	\label{fig:weight_vis}
\end{figure*}
\textit{2) Visual Demonstration of Various Intervention Values on the Inferred Images:} In Fig. \ref{fig:weight_vis}, we examine the influence of the intervention factor $\alpha$ specified in Equation (\ref{weighting_equation}) on the images inferred under different scenarios involving the transmission of various types of semantic units. Similarly, the units from the first column are transmitted to the receiver to commence the image generation process. Following a 20-step denoising period, the units from the second column are subsequently transmitted. We compare the reconstruction results using sequential conditional denoising with intervention factors ranging from $2$ to $20$ against representative failure cases produced by the traditional denoising method (shown in the column labeled $\alpha = 0$). Take the transmission scenario in the first row as an example. When $\alpha = 0$, the inferred image resembles an outdoor scene, with wild grass on either side of the dog and the object beneath the dog appearing more like wooden steps. With the increase of $\alpha$, the extended wooden steps transform into a round, enclosed table, and a chair commonly associated with the table appears in the background. Even the wild grass on either side of the dog gradually turns into patterns on the table. 
However, it is important to note that these visual results also demonstrate that a higher $\alpha$ is not always better. We observe that when $\alpha$ is very high, the semantic information related to previously received units is diminished in the inferred images. For example, at $\alpha = 20$, a strange pattern appears on the dog's nose in the first column, and the aesthetics of the woman's image in the second column are also reduced. Therefore, in practice, it is necessary to design an appropriate intervention value to balance the influence of previously received units and newly received units on the final inferred image.  

\subsection{Evaluation of the Semantic-aware Resource Allocation Module}

\begin{figure} 
	\centering
	\includegraphics[width=8cm]{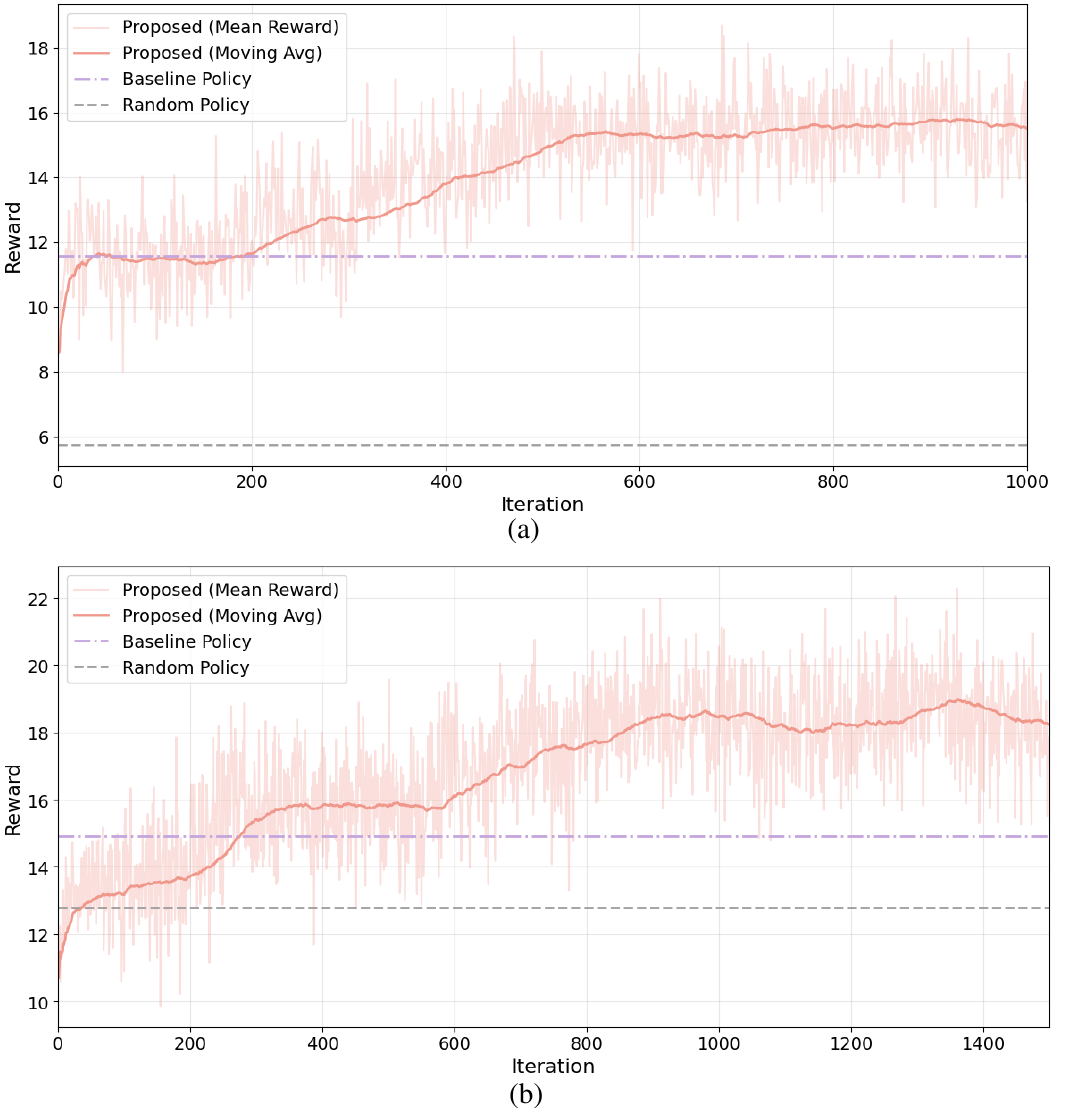}
	\caption{Validation reward relative to the training iterations. In (a), the available bandwidth is 10 KHz. In (b), the available bandwidth is 20 KHz.}
	\label{fig:SRA}
\end{figure}

In this section, we evaluate the performance of the semantic-aware resource allocation module. The proposed method is compared against the following two benchmark policies: 1) Random Policy: The available bandwidth is randomly allocated among all extracted semantic units. 2) Baseline Policy: The limited bandwidth is allocated based on user task importance, prioritizing users with higher task importance. The results are shown in Fig. \ref{fig:SRA} (a) and (b), where the available bandwidth is $10$ KHz and $20$ KHz, respectively. In Fig. \ref{fig:SRA} (a), the proposed method's mean reward rises until converging around episode 580, stabilizing between 15–16. In contrast, the Baseline Policy remains below 12, and the Random Policy performs worst, staying below 6. In Fig. \ref{fig:SRA} (b), as the bandwidth constraint is relaxed, the mean reward attained by the proposed method continues to increase until convergence occurs at approximately 1000 episodes, stabilizing at a value between 18 and 19. Meanwhile, the mean reward of the Baseline Policy is around 15, whereas the Random Policy remains consistently low at around 13.

Across both Fig. \ref{fig:SRA} (a) and (b), the proposed method consistently outperforms both the Baseline Policy and the Random Policy in terms of absolute reward value. This result demonstrates its strong adaptability to channel state variations and its overall superiority in optimizing resource allocation.

\subsection{Overall Analysis of Combined Refinements}
In this section, we examine whether the proposed three modules can function cohesively to improve overall system performance. Initially, we explore the impact of various intervention factors of the sequential conditional denoising module to find a more suitable one for use in the FAST framework. Subsequently, we present numerical results comparing the system precision, latency, and efficiency of the proposed FAST framework against those of conventional GSC systems.

\begin{figure} 
	\centering
	\includegraphics[width=8cm]{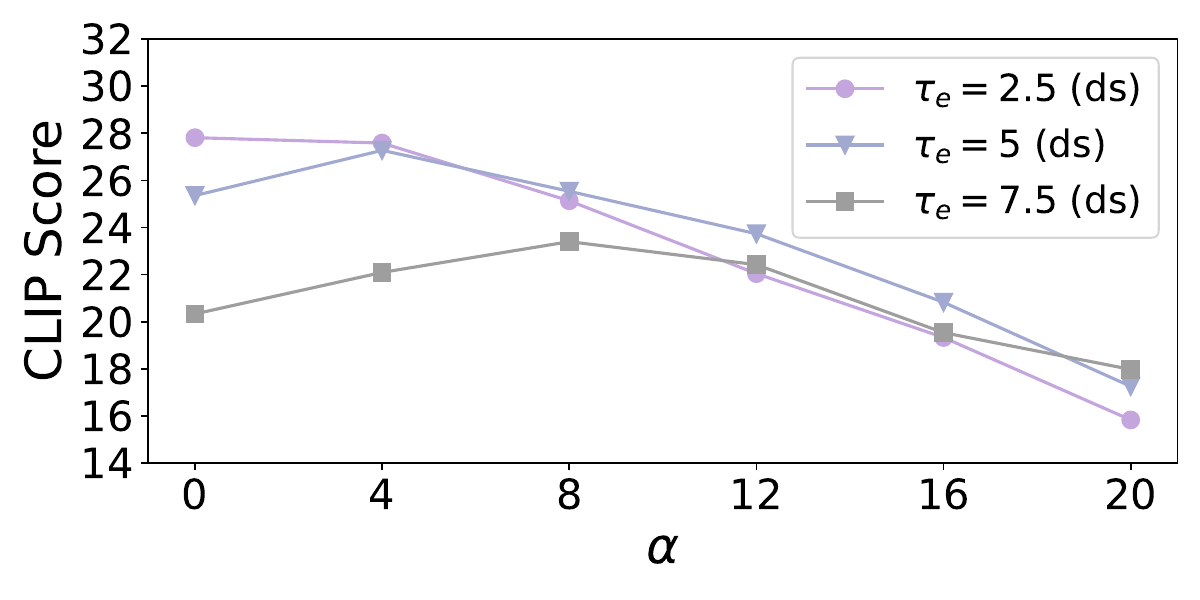}
	\caption{Variations in CLIP scores relative to the intervention factor under different latencies for extracting a single semantic unit, with extraction latency measured in denoising steps (ds).}
	\label{fig:weight_plot}
\end{figure}

\textit{1) Numerical Results of Various Intervention Factor Values on the System Performance:} Herein, we examine the variations in CLIP scores with respect to the intervention factor $\alpha$, aiming to identify a more suitable value to work within the FAST under our default and two alternative extraction latency settings. To synchronize with the training process of the temporal prompt engineering method, we segment every 5 denoising steps for the alternative setting where the latency for extracting a single semantic unit ${\tau _e}$ equals the duration of 2.5 denoising steps, every 10 denoising steps for the default setting where the latency for extracting a single semantic unit ${\tau _e}$ equals the duration of 5 denoising steps, and every 15 denoising steps for the alternative setting where the latency for extracting a single semantic unit ${\tau _e}$ equals the duration of 7.5 denoising steps. As illustrated in Fig. \ref{fig:weight_plot}, under the same denoising speed, a higher intervention value is desired for systems with longer extraction latency. This is due to the fact that, once the initial semantic units initiate the denoising process, the extended extraction latency prolongs the incorporation time for each subsequent semantic unit. Consequently, the residual noise level in intermediate results is lower compared to settings with shorter extraction latency. In such scenarios, a greater noise space intervention capability is necessary to adjust the noise space effectively and incorporate the newly arrived units under conditions of extended extraction latency. 

\begin{figure*}
	\centering
	\includegraphics[width=18cm]{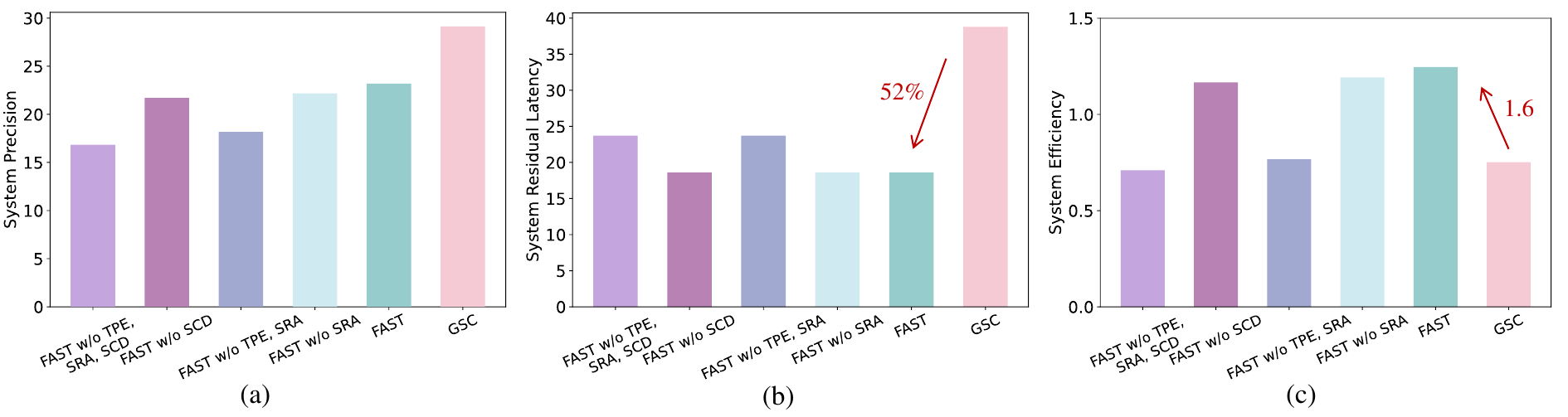}
	\caption{Comparison of numerical results for the conventional GSC scheme, the full FAST framework, and its variants incorporating temporal prompt engineering (TPE), semantic-aware resource allocation (SRA), and sequential conditional denoising (SCD) modules. (a) System precision: the weighted average of individual reconstruction precision; (b) Residual system latency: the weighted average of individual residual latency, which is defined as the latency exceeding the fixed reconstruction duration and measured in denoising steps; (c) System efficiency: the weighted average of individual efficiency, which is defined as the ratio of a user's precision score to their residual latency.}
	\label{fig:overall}
\end{figure*}

\textit{2) Numerical Results of System Performance:} Based on the numerical results assessing the impact of various intervention factor values on the reconstruction precision, we set the intervention factor $\alpha$ at 4. To comprehensively evaluate the proposed framework, we define three performance metrics for each user: precision, residual latency, and efficiency. Precision quantifies reconstruction fidelity and is measured by the CLIP score. Residual latency is the latency exceeding the fixed reconstruction duration, measured in denoising steps. Efficiency is defined as the ratio of a user's precision to their residual latency. The corresponding system-level metrics are then derived as the weighted average of individual user metrics, weighted by each user's task importance score. Finally, a comprehensive evaluation of the conventional GSC scheme, FAST variants, and the full FAST framework is presented in Fig. \ref{fig:overall}. Numerical results demonstrate that integrating the temporal prompt engineering and semantic-aware resource allocation modules into the FAST framework collectively reduces residual system latency by 21\% and improves system precision by 29\%, compared to a baseline without these components. Furthermore, the sequential conditional denoising module contributes an additional 8\% gain in system precision. In terms of overall efficiency, the FAST framework achieves a metric approximately 1.6 times higher than that of conventional GSC. These findings confirm that the proposed architecture exhibits enhanced potential for deployment in multi-user GSC scenarios characterized by stringent communication and computational constraints.

\section{Conclusion}\label{Conclusion}
This paper has introduced FAST, a flexible and adaptive semantic transmission framework designed to address the challenges of complex applications such as digital twin systems and virtual reality environments. To overcome the computational constraints in such environments, we have developed a novel sequential semantic extraction approach guided by temporal prompt engineering, enabling efficient distillation and transmission of semantic units. At the receiver side, a sequential conditional denoising module has been implemented to progressively reconstruct content using the streamed semantic information. For multi-user scenarios, we have created a semantic-aware resource allocation scheme that dynamically distributes bandwidth through joint optimization of semantic dependencies, task priorities, and channel states. Experimental results have confirmed that the proposed framework has maintained reconstruction precision comparable to conventional GSC systems while achieving substantial improvements in both latency reduction and overall system efficiency. These findings have demonstrated its enhanced potential for deployment in multi-user GSC scenarios characterized by stringent communication and computational constraints.

Looking forward, several research directions merit further investigation. First, extending the FAST framework to support more diverse generative models beyond diffusion models represents a promising avenue. Second, exploring cross-modal semantic alignment techniques could enhance the framework's capability in handling increasingly heterogeneous data sources. Finally, investigating security and privacy preservation mechanisms for semantic transmission in sensitive applications remains an important challenge for future work.

\bibliographystyle{IEEEtran}
\bibliography{IEEEabrv,myre}

\end{document}